\documentclass[10pt]{article}
\usepackage{jheppub}
\usepackage{enumerate,epsfig,amsmath,capt-of,wasysym,mathrsfs,graphicx,amsfonts,amsbsy,amscd,multirow,rotating,slashed, latexsym, verbatim, cancel} 
\usepackage[utf8]{inputenc}
\usepackage {multirow}
\usepackage[normalem]{ulem}
\usepackage{color}
\definecolor{myred}{rgb}{0.6,0,0} 
\definecolor{myblue}{rgb}{0,0.2,0.4}
\definecolor{mygreen}{rgb}{0,0.9,0.1}
\definecolor{hc}{rgb}{.9,0.1,0.7}
\definecolor{hcout}{rgb}{.9,0.7,0.9}
\definecolor{Orange}{rgb}{1.,0.65,0.}
\newcommand{\lsim}{{\;\raise0.3ex\hbox{$<$\kern-0.75em\raise-1.1ex\hbox{$\sim$}}\;}}
\newcommand{\gsim}{{\;\raise0.3ex\hbox{$>$\kern-0.75em\raise-1.1ex\hbox{$\sim$}}\;}}
\usepackage{array}
\usepackage{bm} 
\usepackage{cleveref}
\newcommand{\be}{\begin{equation}}
\newcommand{\ee}{\end{equation}}
\newcommand{\bes}{\begin{equation*}}
\newcommand{\ees}{\end{equation*}}
\newcommand{\bea}{\begin{eqnarray}}
\newcommand{\eea}{\end{eqnarray}}
\newcommand{\beas}{\begin{eqnarray*}}
\newcommand{\eeas}{\end{eqnarray*}}
\newcommand{\la}{\lambda}
\newcommand{\wt}{\widetilde}
\title{Leptophilic bosons and muon g-2 at lepton colliders}
\author{Eung Jin Chun,}
\author{Tanmoy Mondal} 
\affiliation{School of Physics, Korea Institute for Advanced Study,\\ Seoul 02455, Republic of Korea }
\emailAdd{ejchun@kias.re.kr}
\emailAdd{tanmoy@kias.re.kr}
  
\abstract{
  A light leptophilic boson (scalar or pseudoscalar)  has been postulated to explain the muon g-2 anomaly and 
  could be a portal to dark matter. Realizing the leptophilic nature of a singlet boson in the framework of 
  the two-Higgs-doublet-Model of type-X, we identify the parameter space viable for the explanation of the 
  updated muon  g-2 discrepancy.  It is then shown that such a hypothetical particle will be unambiguously 
  ruled out or discovered via the Yukawa process at a lepton collider designed as a Higgs factory.
}

\preprint{KIAS-P21014}
\date{\today}

\begin{document}
\maketitle
\section{Introduction}

There have been extensive studies on new physics below the weak scale which can be looked for at various 
high-energy and high-intensity experiments in vast ranges of new particle masses and couplings~\cite{Beacham:2019nyx}.
One of the interesting possibilities is the presence of a light spin-0 particle mixing with the Higgs boson
\cite{Kim:2008pp,Batell:2009jf,Bezrukov:2009yw,Clarke:2013aya} which could be a portal to dark matter, or an inflaton.
A scalar particle mixing with the SM Higgs boson interacts with the fermions with a coupling strength proportional to 
the fermion mass. Such a particle is severely constrained due to its hadronic interactions. 
On the other hand, if the scalar mediator couples predominantly to the leptons \cite{Batell:2016ove}, 
substantial parameter space still remains open and viable to accommodate the $(g-2)_\mu$ anomaly 
which is confirmed by FNAL~\cite{Bennett:2006fi, fnal21,Aoyama:2020ynm}. 
  In connection to the dark sector, non-observation of any signal in the direct detection experiments may indicate a
  leptophilic mediator for dark matter. If the mediator is a pseudoscalar, the direct detection of dark matter becomes 
  out of question since its interaction with matter is velocity suppressed \cite{Dolan:2014ska}.

\medskip

A two-Higgs-doublet Model (2HDM) can provide a natural framework to realize the leptophilic nature of a light  
scalar or pseudoscalar. Phenomenology of scalar or pseudoscalar extension of  2HDM has been studied mainly in 
the context of dark matter~\cite{Ipek:2014gua,No:2015xqa,Bell:2016ekl,Goncalves:2016iyg,Bauer:2017ota,Bell:2017rgi,Arcadi:2017wqi,Sanderson:2018lmj,Arhrib:2018qmw,Arcadi:2018pfo,Abe:2019wjw,Arcadi:2020gge,Butterworth:2020vnb,Chen:2021rnl}. 
A typical LHC search for a light leptophilic boson relies on the decay of the SM Higgs boson to a  pair of light 
particles which subsequently decay into leptons. This could be an important discovery channel if it is allowed 
kinetically and  the corresponding coupling is sizable.  More direct searches can be performed through the Yukawa 
production followed by leptonic decays. The new results from BABAR Collaboration~\cite{BABAR:2020oaf} have excluded  a leptophilic scalar between 0.1 MeV to a few GeV in the coupling range which can explain the $(g-2)_\mu$ anomaly. 
Incorporating the one- and two-loop contributions properly,  we will see that the leptophilic scalar mass window  of a few to 10 GeV is still open to accommodate the muon $(g-2)$ anomaly. 
The other region below the BABAR limit has been ruled out by the electron beam dump experiments~\cite{Bjorken:1988as,Davier:1989wz}.

The situation for a leptophilic pseudoscalar is completely different. 
First of all, the minimal 2HDM of type-X can accommodate the muon $(g-2)$ anomaly for large $\tan\beta$ and a CP-odd Higgs boson above 10 GeV~\cite{Cao:2009as,Broggio:2014mna,Wang:2014sda,Abe:2015oca,Chun:2016hzs}\footnote{For some recent developments along this direction see~\cite{Han:2018znu,Wang:2019ngf,Sabatta:2019nfg,Jana:2020pxx,Jana:2020joi,Chen:2021jok}. A comprehensive discussion about BSM explanation of $(g-2)_\mu$ in the light of FNAL result can be found in~\cite{Athron:2021iuf}.}.
Considering a light pseudoscalar mixing with a heavier CP-odd Higgs boson, the muon $(g-2)$ favored region is extended to lower mass which is however above the BABAR reach, and there is no direct search limit so far. 

\medskip

A light particle at the GeV scale is in the border zone between high-intensity and high-energy searches, and hence it is challenging to extend the search limits on either side.
The purpose of this paper is to see how far in the lower mass range  future collider experiments can reach and 
check whether  the muon $(g-2)$ region of a scalar and pseudoscalar can be covered.  
At a lepton collider,  a leptophilic boson can be produced via  the Yukawa process~\cite{Kalinowski:1988sp,Kalinowski:1995dm,Kalinowski:1996nr,Dawson:1998qq,Grzadkowski:1999ye,Abbiendi:2001kp,Chun:2019sjo} where a tau lepton radiates a boson, and then decays almost exclusively to a pair of leptons. 
If allowed kinematically, it leads to the conventional signal of four tau-lepton final states. 
It is possible to probe the four-tau final state and use collinear approximation to reconstruct the mass at the ILC~\cite{Chun:2019sjo}. However, for a lighter boson,
such a search method is not feasible as the decay products  will be too collimated to be distinguished as 
two separate tau-tagged jets.

 In this paper, we propose to search for a muon in association with an 
opposite sign charged track as the signal for a light leptophilic boson. 
The charged track will be adjacent to the muon 
since both of them originate from the decay of a collimated pair of tau leptons coming from the boson. 
Precision measurement of the Higgs boson is the prime target for any future lepton collider which will run as a `Higgs factory' with a centre-of-mass energy of 250 GeV. 
We will consider ILC as an example and show that the proposed signal efficiently probes a light 
$\phi$ in the mass and coupling range beyond the BABAR limit. 
Our signature will be a complementary exploration of any spin-0 particle in the GeV scale mass range. 
It is worthwhile to emphasize that the signals at the lepton collider under the consideration arise
solely from the Yukawa interaction of a boson with leptons, and thus are decisive to confirm or disprove the 
presence of a leptophilic boson.

\section{A Framework for Leptophilic Scalar and Pseudoscalar}\label{sec:model}
We will briefly discuss the scalar and pseudoscalar extension of the 2HDM scenario~\cite{Gunion:1989we,Djouadi:2005gj,Branco:2011iw}  as a framework for an extra leptophilic boson.
 The scalar potential of 2HDM is given by,
\begin{eqnarray}
\nonumber V_{\mathrm{2HDM}} &=& -m_{11}^2\Phi_1^{\dagger}\Phi_1 - m_{22}^2\Phi_2^{\dagger}\Phi_2 -m_{12}^2\Big[\Phi_1^{\dagger}\Phi_2 + \Phi_2^{\dagger}\Phi_1\Big]
+\frac{1}{2}\lambda_1\left(\Phi_1^\dagger\Phi_1\right)^2+\frac{1}{2}\lambda_2\left(\Phi_2^\dagger\Phi_2\right)^2 \\
\nonumber && +\lambda_3\left(\Phi_1^\dagger\Phi_1\right)\left(\Phi_2^\dagger\Phi_2\right)+\lambda_4\left(\Phi_1^\dagger\Phi_2\right)\left(\Phi_2^\dagger\Phi_1\right)
+\frac{1}{2}\lambda_5\Big\{ \left(\Phi_1^\dagger\Phi_2\right)^2+ \left(\Phi_2^\dagger\Phi_1\right)^2\Big\}.
\label{eq:2hdm-pot}
\end{eqnarray}
The dimensionful coupling $m_{12}^2$ breaks the $\mathbb{Z}_2$ charge introduced to forbid the flavor changing neutral current interactions. 
Since we are interested in leptophilic extensions, we will consider the type-X model where $\Phi_1$ and the right-handed leptons are odd under the $\mathbb{Z}_2$ symmetry, whereas all other particles are even under the same symmetry. Consequently, the Yukawa Lagrangian is given by,
\begin{equation}\label{eq:yukawa}
-{\cal L}_Y= Y^u\bar{ Q_L} \wt \Phi_2 u_R + Y^d  \bar{ Q_L} \Phi_2 d_R+Y^e\bar{ l_L} \Phi_1 e_R + h.c.,
\end{equation}
where $\wt \Phi_2=i\sigma_2\Phi_2^*$. The scalar doublets can be parameterized as,
$\Phi_i = \left(\phi_i^+,\left(v_i+\Phi_i^{0R}+i\, \Phi_i^{0I}\right)/\sqrt{2}\right)^T$ and as usual $\tan\beta=v_2/v_1$ and $v=\sqrt{v_1^2+v_2^2}=246$ GeV.

\subsection{Scalar Extension of 2HDM} 
In this scenario a CP even scalar ($S$) is added to the scalar sector of 2HDM.  The scalar potential mixing $S$ with the doublets $\Phi_{1,2}$ is given by,
\begin{eqnarray}
 V_{\mathrm{S}-\Phi} =  A_{12}\Big[\Phi_1^{\dagger}\Phi_2 + \Phi_2^{\dagger}\Phi_1\Big] S + A_{11} \Phi_1^{\dagger}\Phi_1 S + A_{22} \Phi_2^{\dagger}\Phi_2 S  + \dfrac{\la_{1S}}{2} \Phi_1^{\dagger}\Phi_1 S^2 +\dfrac{\la_{2S}}{2} \Phi_2^{\dagger}\Phi_2 S^2
\end{eqnarray}
After the spontaneous electroweak symmetry breaking (EWSB) the scalar sector is comprised of on charged Higgs, one CP-odd neutral scalar and three CP-even  neutral scalars.
The rotation matrix which diagonalize the $3\times3$ neutral CP-even scalar mass matrix will mix the gauge eigenstates,
\be\label{eq:scalar-mix}
\begin{pmatrix}
 \Phi_1^{0R}\\\Phi_2^{0R}\\S
\end{pmatrix}
\approx\begin{pmatrix}
  -s_\alpha&c_\alpha&s_{\theta_1}\\
    c_\alpha&s_\alpha&s_{\theta_2}\\
     -s_{\theta_1}& -s_{\theta_2}&1\\
 \end{pmatrix}
 \begin{pmatrix}
  h_{125}\\H \\ \phi
 \end{pmatrix}.
\ee
for small $\sin\theta_{1,2}$.
%
Here $H$ is heavy CP-even scalar which we assume to be degenerate with the charged scalar ($H^\pm$) and the pseudoscalar ($A$) in our analysis.
The mass eigenstates $\phi$ is supposed to be light to explain the muon $g-2$ anomaly. 

After symmetry breaking we can write the interaction of the light scalar $\phi$ with the fermion mass eigenstates using Eq.~\ref{eq:yukawa} and \ref{eq:scalar-mix},

\be\label{eq:yuk-H_2}
\mathcal L_{\mathrm{Yukawa}}^{\mathrm{Physical}} \supset
- \frac{m_q}{v}~ \dfrac{\sin\theta_2}{ \sin\beta}~  \phi \overline{q} q -  \frac{m_\ell}{v}~ \dfrac{\sin\theta_1}{\cos\beta} ~\phi \overline{\ell} \ell
\ee
where $q=u,d$.
From the above equation it is evident that  $\phi$ will have a large coupling to leptons for large $\tan\beta$. 
The angle $\theta_2$ dictates the mixing of $\phi$ with $\Phi_2$ which is mostly the SM Higgs doublet near the alignment limit. 
By keeping this mixing small, it is possible to evade present bounds on light scalar coming from LHC and the proton beam dump experiments. The effective lepton Yukawa coupling of $\phi$ will be denoted by
\be\label{eq:2hdms-eff-Yuk}
\xi_{\phi}^\ell =  \dfrac{\sin\theta_1}{\cos\beta} \approx \tan\beta \sin\theta_1
\ee
for large $\tan\beta$.

\subsection{Pseudoscalar Extension of the 2HDM}
In this case a CP-odd scalar $P$ is added to the 2HDM sector.
The additional scalar potential mixing $P$ with $\Phi$ reads as,
\begin{eqnarray}
 V_{\mathrm{P}-\Phi} =  i\,B_{12}\Big[\Phi_1^{\dagger}\Phi_2 - \Phi_2^{\dagger}\Phi_1\Big] P + \dfrac{\la_{1P}}{2} \Phi_1^{\dagger}\Phi_1 P^2 +\dfrac{\la_{2P}}{2} \Phi_2^{\dagger}\Phi_2 P^2 
\end{eqnarray}
After the EWSB the imaginary components of the doublets $\Phi_i^{0I}$ mixes to form the Goldstone of the $Z$ boson and $\hat A (=-s_\beta\Phi_1^{0I}+c_\beta\Phi_2^{0I})$. Then $\hat A$ mixes with $P$ to generate two mass eigenstates,
\be\label{eq:pseudoscalar-mixing}
\begin{pmatrix}
 \hat A\\P
\end{pmatrix}
=\begin{pmatrix}
  c_\theta&-s_\theta\\
 s_\theta&c_\theta
 \end{pmatrix}
 \begin{pmatrix}
 A\\ \phi
 \end{pmatrix}
\ee
 where the mixing angle can be written as,
 \be
 \tan2\theta = \dfrac{2B_{12}v}{m_A^2-m_\phi^2}.
 \ee
 Note that we used the same notation $\phi$ for the scalar and pseudoscalar mass eigenstates.
 For our analysis, we will take the same mass for the heavy Higgs bosons of 2HDM:  $A$, $H$ and $H^\pm$.
The light mass eigenstate $\phi$ will interact with fermions through the mixing as follows:
\be\label{eq:yuk-light-a}
\mathcal{L}\supset  \dfrac{\sin\theta}{\tan\beta} \dfrac{m_u}{v}\,   i \phi  \bar u \gamma^5 u +  \dfrac{(-\sin\theta)}{\tan\beta} \dfrac{m_d}{v} \, i\phi \bar d \gamma^5 d +\tan\beta\sin\theta \dfrac{m_\ell}{v}\,  i \phi \bar \ell \gamma^5 \ell.
\ee
Again we define  the effective lepton coupling as 
\be\label{eq:2hdma-eff-Yuk}
 \xi_\phi^\ell =\tan\beta\sin\theta
\ee
and the corresponding quark coupling becomes $\xi_\phi^q \approx \xi_\phi^l/\tan^2\beta$ which is highly suppressed at large $\tan\beta$.

\subsection{Fit to the muon $g-2$ and lepton universality constraints}

Extending the analysis in \cite{Chun:2016hzs} to the case of the scalar/pseudoscalar mixing with 2HDM, 
one can identify the parameter region of $(m_\phi, \xi^l_\phi)$ consistent with the new result on 
$a_\mu = (g-2)_\mu/2$ \cite{fnal21}:
\begin{equation}
a_\mu(\mbox{Exp})-a_\mu(\mbox{SM})=(251\pm59) \times 10^{-11}.
\end{equation}
Our 2$\sigma$ favored region does not change too much as there are a little shifts in both the  central value and the error. 
We also plot the 2$\sigma$ bounds from the lepton universality conditions in the lepton
(tau) decays and the leptonic $Z$ decays ($Z\to ll$).  For this, we take $\xi^l_\phi =\tan\beta \sin\theta \lsim 100$, which corresponds to the tau coupling ${m_\tau \over v} \tan\beta$ smaller than $\sim 1$.
The heavy Higgs bosons are taken to be degenerate at 300 GeV and thus their contribution to the muon $(g-2)$ is negligible.

\section{Search at Higgs Factory}\label{sec:simulation}
 \begin{figure}[!t]
 \begin{center}
  \includegraphics[width=15cm]{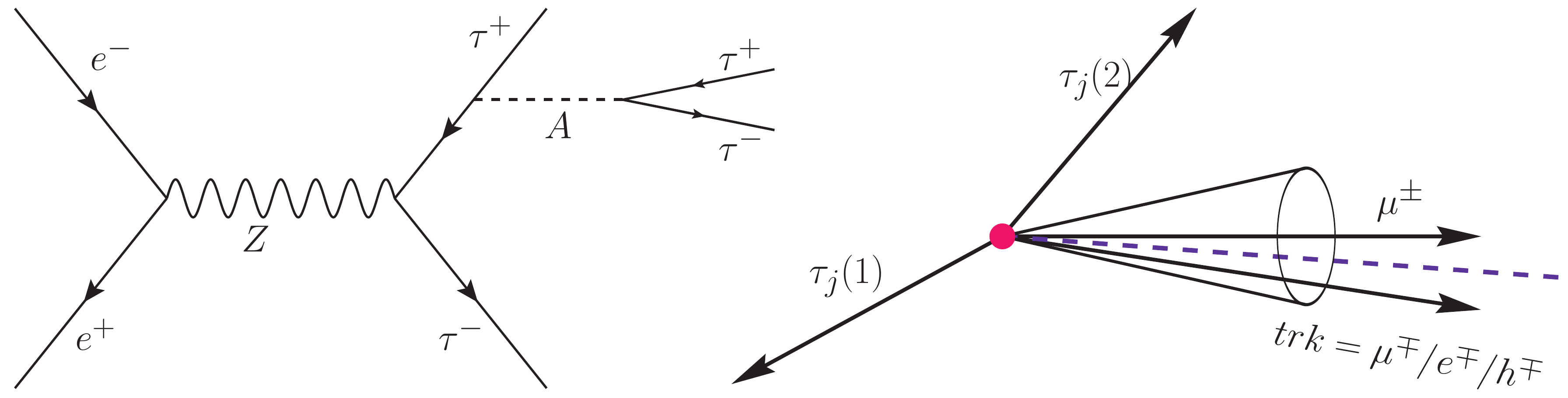}
  \caption{Signal topology of a light boson production at the lepton collider. The dashed line in the right panel depicts the direction 
  of light boson $\phi$.}
  \label{fig:signal-topology}
 \end{center}
 \end{figure}
 
The Yukawa production of a leptophilic scalar or pseudoscalar is given by
\be
e^+ e^- \to Z^*/\gamma^* \to \tau^+ \tau^- \phi.\nonumber 
\ee
When $m_\phi > 2 m_\tau$, which is the case under consideration, $\phi$ decays dominantly to a pair of taus. 
For a light $\phi$, its decay products will be collimated, and it is not feasible to isolate four final state tau leptons. 
To overcome this difficulty, we propose to look for a muon and a single charge track in the vicinity of the muon.
In this case, one of the tau coming from $\phi$ will decay to a muon, and the other  will give rise to a charged track. 
The charge track can originate from the leptonic decay mode or hadronic one-prong decay of the tau lepton, 
thus includes 85\% of the tau decay products.
In addition to the muon and track system, there will be two additional tau-tagged jets.  
The signal topology is shown in Fig.~\ref{fig:signal-topology}.
The dominant background comes from di-boson production, 
\be
e^+ e^-\to Z/\gamma^*\;Z/\gamma^* \to 2\mu\; 2\tau ,\; 4\tau. 
\ee
There is also a subdominant contribution from the Higgs-strahlung process.

\subsection{Event Simulation and Signal Selection}
 We use $\texttt{MadGraph5\_aMC@NLO}$ \cite{Alwall:2011uj,Alwall:2014hca}  to generate 
 parton level signal and background events and \texttt{PYTHIA8} \cite{Sjostrand:2006za,Sjostrand:2014zea} for the subsequent decay, showering and hadronization. For our  analysis  we have polarized beam with $P(e^+,e^-)=(+30\%,-80\%)$~\cite{Behnke:2013lya}. 
 The $\tau$ decays are incorporated via \texttt{TAUOLA}  \cite{Jadach:1993hs} integrated in $\texttt{MadGraph5\_aMC@NLO}$. 
 For detector effects we use \texttt{Delphes3} \cite{deFavereau:2013fsa} with the ILD detector  card to simulate the detector effects. 
 Jets are clustered using the  longitudinal-kT algorithm~\cite{Catani:1993hr,Ellis:1993tq} with $R = 0.4$. In \texttt{Delphes3},  we assumed 
 the tagging efficiency of $\tau$ jets $\epsilon_\tau = 90\%$~\cite{Jeans:2018anq}. Since we are interested a charge track close to the muon 
 we remove the muon isolation criteria. We imposed the \textit{pre-selection criteria} that all the tau tagged and the muon should have minimum energy 
 of 20 GeV and should have $|\eta| <2.3$  which corresponds to $|\cos\theta| < 0.98$. In addition we require that both the tau-tagged jets and the muon 
 should be well separated with $\Delta R\geq 0.4$. 
 
\begin{figure}[!t]
 \begin{center}
  \includegraphics[width=8.8cm]{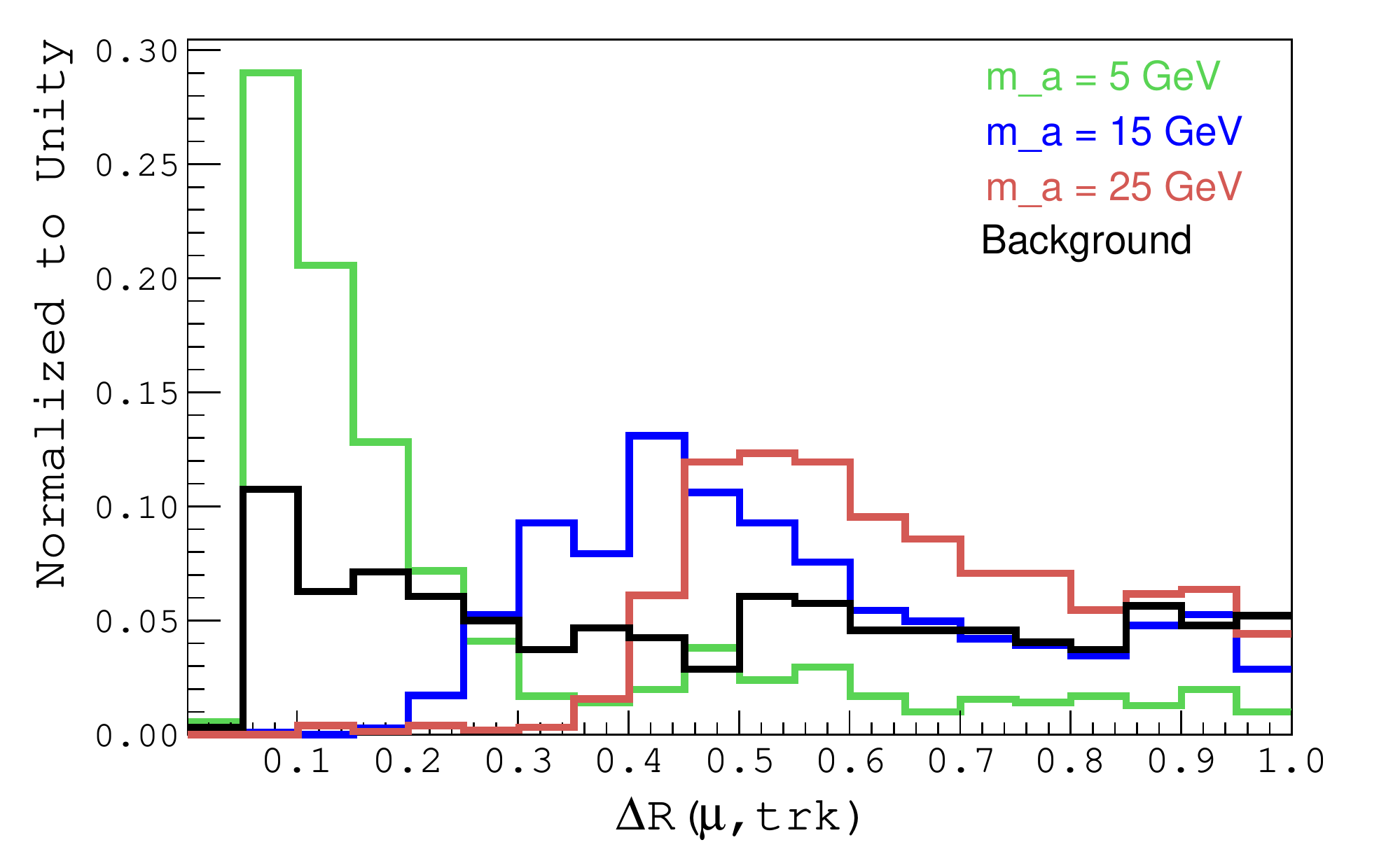}
  \includegraphics[width=6.2cm]{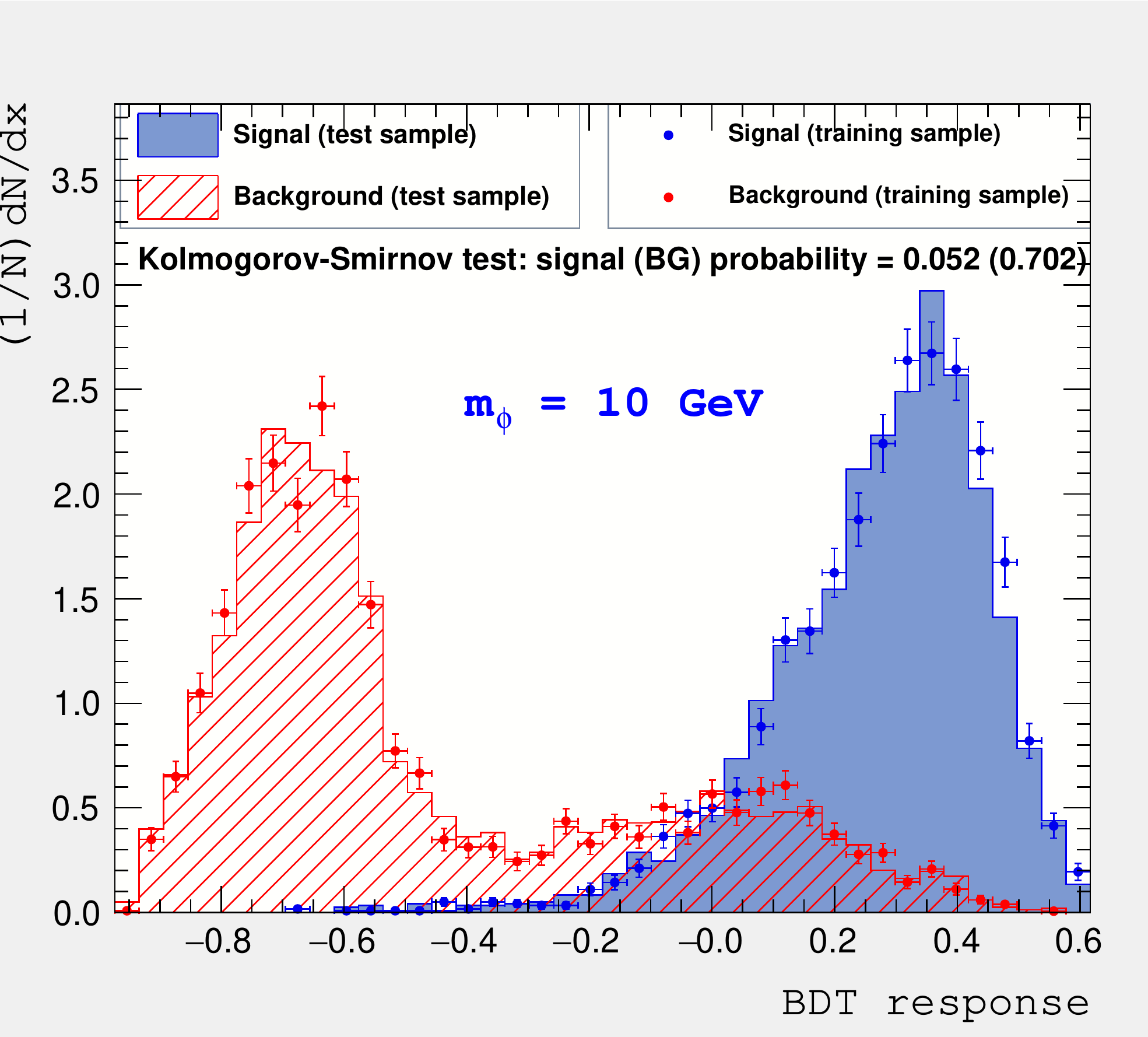}
  \caption{Angular separation between muon and a single charge track for different values of $m_\phi$. 
  The black histogram shows the background distribution. Evidently the separation is small for light $\phi$ and increases as $m_\phi$ increases.\emph{Right Panel:} Here we show BDT response of signal and background for $m_\phi=10$ GeV.}
    \label{fig:deltaRmutrk}
 \end{center}
 \end{figure}

 There should be a single opposite sign charge track  with  $p_T > 2.5$ GeV and the angular distance between the muon and the charge track 
 $\Delta R(\mu^\pm,\mathtt{trk^\mp})$ should be less than 1.0. In Fig.~\ref{fig:deltaRmutrk} we show the variation of $\Delta R(\mu^\pm,\mathtt{trk^\mp})$ 
 for different $\phi$ mass and for background processed and as $m_\phi$ increases the angular separation increases.

\subsection{Signal and Background Discrimination}
We have used boosted decision tree (BDT) in the \texttt{TMVA}~\cite{Hocker:2007ht} framework to classify signal and background events. The following seven variable are used as input for the BDT : Energy of muon and the two tagged jets, angular separation between the muon and the tau tagged jets and $\Delta R(\mu^\pm,\mathtt{trk^\mp})$. 
The BDT classifier is shown in right panel of Fig.~\ref{fig:deltaRmutrk} for $m_\phi= $10 GeV. As expected $\Delta R(\mu^\pm,\mathtt{trk^\mp})$ is
the most important variable in the classifier followed by energy of muon.

\section{Results and Discussion}
\begin{figure}
 \begin{center}
   \includegraphics[width=7.5cm]{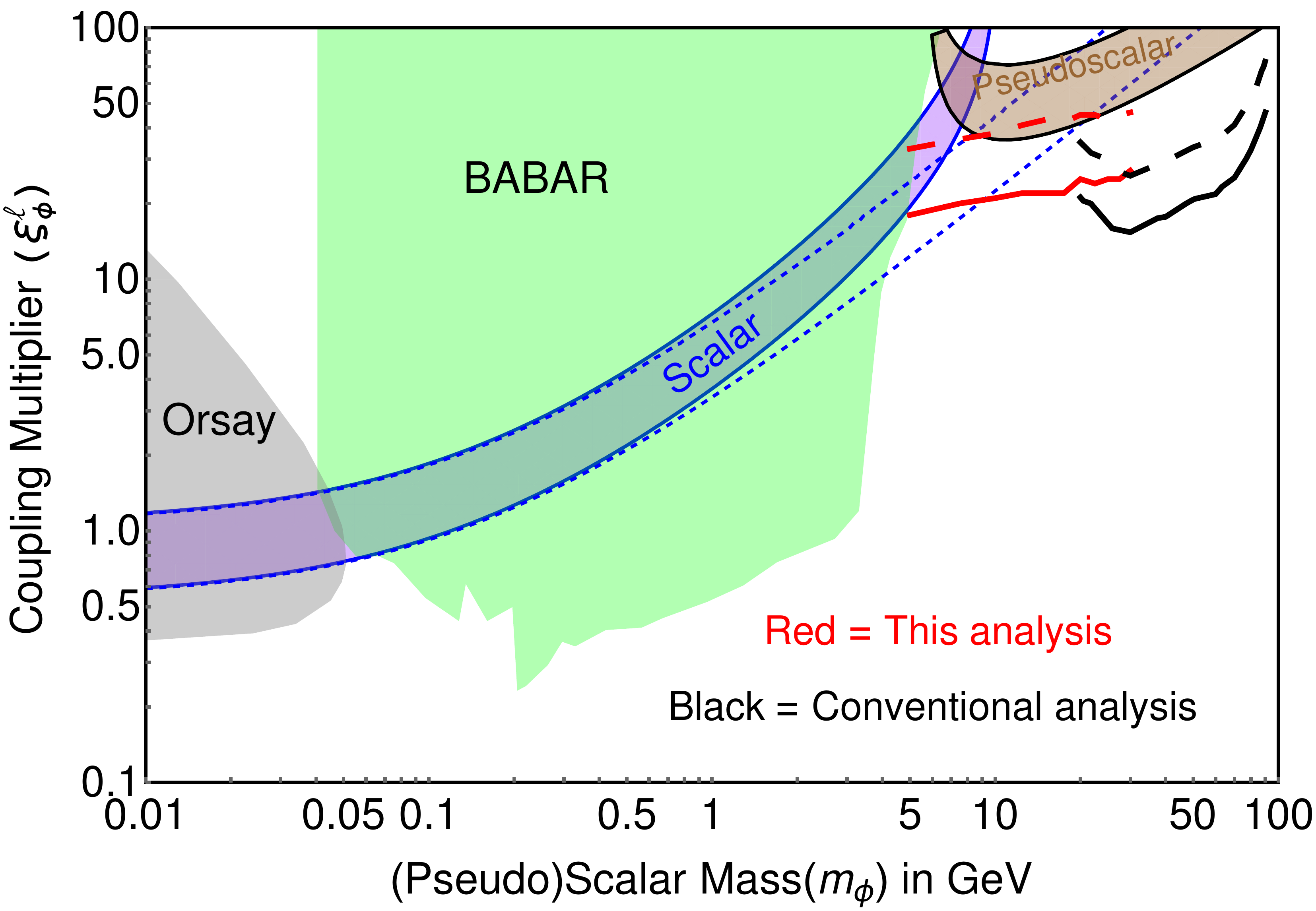}
  \includegraphics[width=7.5cm]{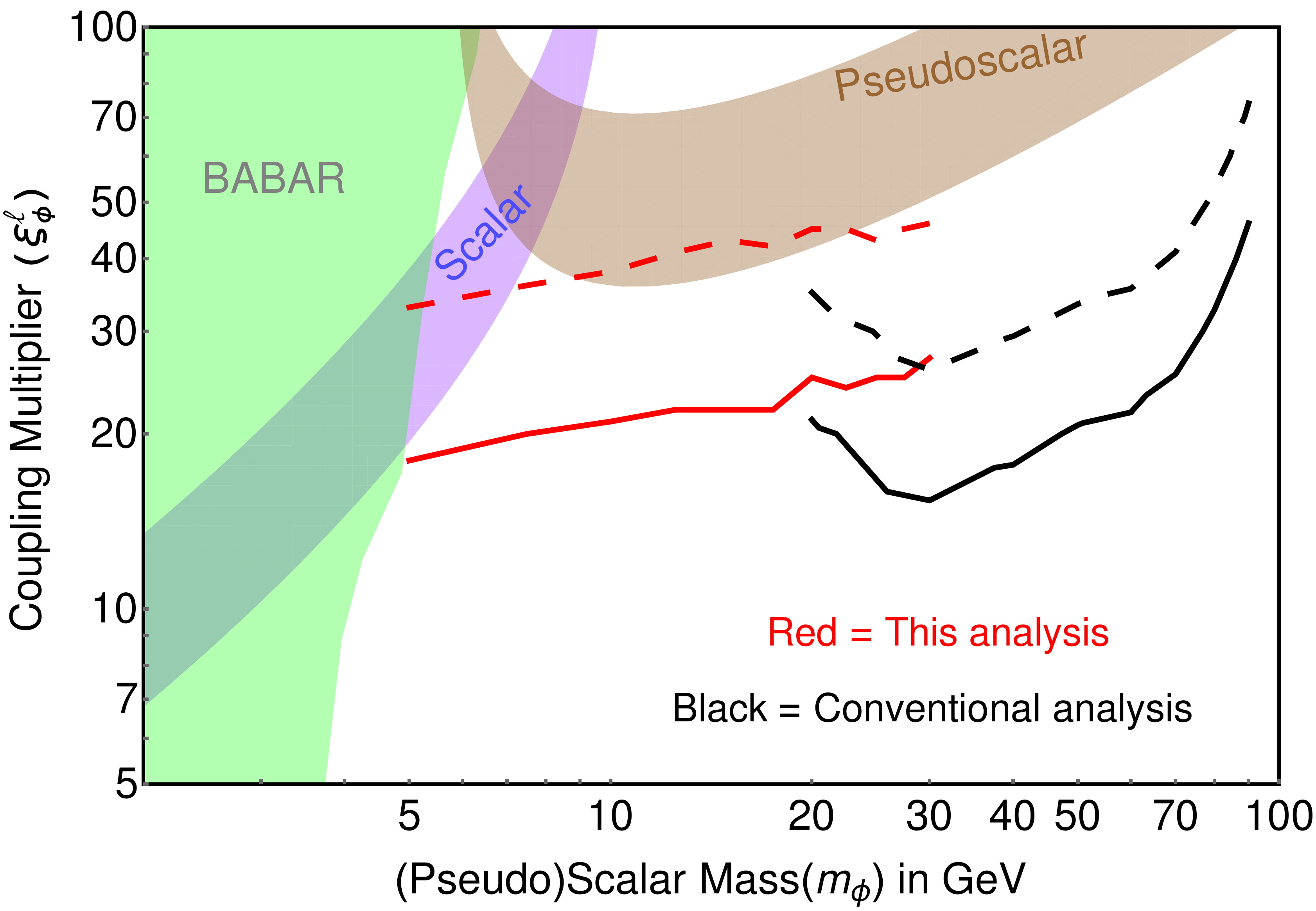}
  \caption{Parameter space of a leptophilic spin-0 particle explaining the muon $(g-2)$ anomaly. Blue (brown) colored region can explain $(g-2)_\mu$ when the spin-0 particle is a scalar (pseudoscalar). The negative two-loop effect in scalar case is apparent for higher mass range where total contribution in inside blue solid region differs from the blue dotted lines.  Substantial part of the parameter space is constrained by the beam dump~\cite{Davier:1989wz} (shown in gray) and BABAR~\cite{BABAR:2020oaf} experiment (shown in green). Limit from ILC is shown in red and black curves where solid and dashed line corresponds to exclusion($2\sigma$) and discovery($5\sigma$) limit at ILC250 with 2$ab^{-1}$ of data. Limit of conventional analysis is taken from ref~\cite{Chun:2019sjo}.}
    \label{fig:result-big-pitcure}
 \end{center}
\end{figure}

We analyze the signal significance with an integrated luminosity of 2000 $fb^{-1}$. We scan the mass of $\phi$ from 5 GeV to 30 GeV with step 2.5 
GeV and for each mass point we employed MVA classifier to obtain signal significance. 

In Fig.~\ref{fig:result-big-pitcure} we have shown our main result in the $m_\phi-\xi^l_\phi$ parameter space. 
The effective lepton coupling $\xi^l_\phi \approx \sin\theta \tan\beta$ is explicitly shown  Eq.~\ref{eq:2hdms-eff-Yuk} for the scalar and in Eq.~\ref{eq:2hdma-eff-Yuk} for the pseudoscalar. The blue (brown) patch in the left panel shows the 
parameter region which can explain the muon $(g-2)$ anomaly for a scalar (pseudoscalar) particle 
in the mass range from 10 MeV to 100 GeV.  
The coupling required for the scalar boson via one loop is shown in the dotted blue lines. Since the 
contribution coming from the two-loop diagram becomes sizable and can cancel the one-loop contribution, 
a larger coupling is necessary when the scalar becomes heavy. The opposite behavior occurs for the
pseudoscalar for which the two-loop contribution needs to dominate over the one-loop contribution.  
Hence, the pseudoscalar has to be  heavier and larger coupling is required. The green and grey areas are 
excluded by BABAR and Orsay experiments. Evidently, $\phi$ has to be heavier than a few GeV to satisfy these constraints. 
Our results on the reach of ILC are shown in the black and red lines.
The red solid and dashed curves display  the 2$\sigma$ exclusion and 
5$\sigma$ discovery limits at ILC with 2$ab^{-1}$ data.
 Since we have used BDT trained for each mass range the limit remains quite strong throughout.
When $m_\phi$ is relatively heavy, the conventional search is possible. For completeness, we have shown the reach of such analysis  in black solid (dashed) line which corresponds to 2$\sigma$ exclusion (5$\sigma$ discovery) region~\cite{Chun:2019sjo}.
 It is evident that ILC is capable of exploring all the still unconstrained viable region.
 In the right panel of Fig.~\ref{fig:result-big-pitcure} we have zoomed into the relevant parameter space.

\begin{figure}
 \begin{center}
  \includegraphics[width=9cm]{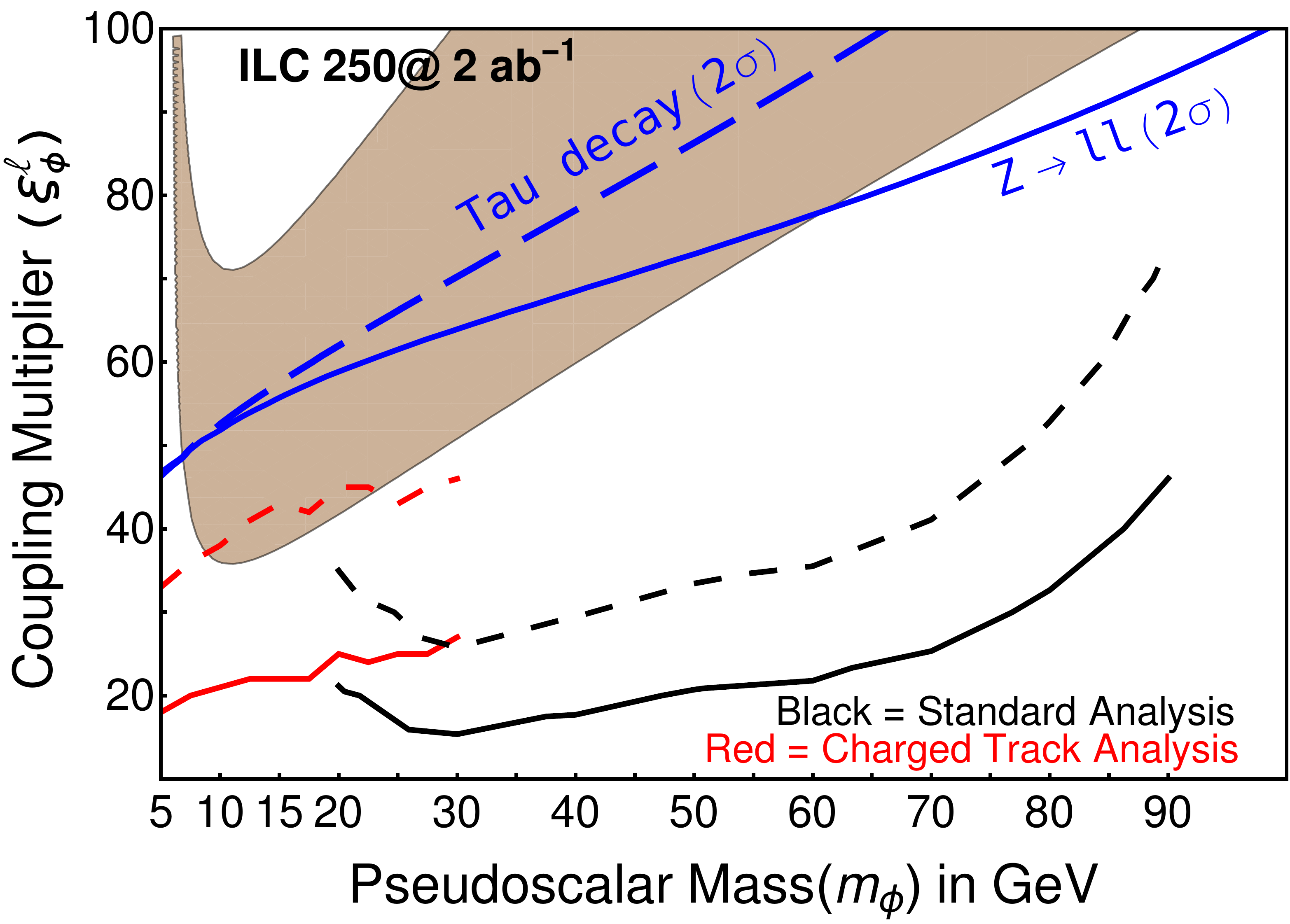}
  \caption{Parameter space in the pseudoscalar extended 2HDM model where the light additional pseudoscalar explain the
$(g-2)_\mu $ anomaly. Strong constraints on the parameter space comes from the lepton universality in the $Z$ and $\tau$ decays. Although substantial parameter space is allowed from lepton universality, ILC will be able to explore the entire parameter space.  Limit from ILC is shown in red and black curves where solid and dashed line corresponds to exclusion($2\sigma$) and discovery($5\sigma$) limit at ILC250 with 2$ab^{-1}$ of data. Limit of conventional analysis is taken from ref~\cite{Chun:2019sjo}.}
    \label{fig:everything-in-one}
 \end{center}
\end{figure}

The parameter space of an additional boson mixing with the 2HDM Higgs bosons gets constrained  
strongly by the lepton universality in the $Z$ and lepton ($\tau$) decays~\cite{Abe:2015oca,Chun:2016hzs}
as shown by the blue solid and dashed lines in Fig.~\ref{fig:everything-in-one}. 
The lepton universality constraint from the $\tau$ decay is sensitive to the value of $\tan\beta$ and we took $\tan\beta = 85$
for the plot.  Here, a large coupling multiplier is already constrained via the lepton universality, and the ILC will be able to explore the remaining parameter space without ambiguity. As before, the red and black solid (dashed) curves show exclusion (discovery) via charged track and conventional analysis respectively.

In leptophilic scenarios, like type-X 2HDM, the coupling of the spin-0 particle with the leptons is proportional 
to the lepton masses, and the tau-Yukawa coupling provides collider signatures for such boson. 
The coupling of the quarks with the spin-0 particle is inversely proportional to $\tan\beta$ and are always small. 
In a more general approach, the aligned 2HDM considers couplings of leptons, up and down type quarks with the 
pseudoscalar is independent of each other.  However, the quark couplings are severely restricted by neutral 
meson mixing, $B_s\to \mu\mu$ and $b\to s\gamma$ observations~\cite{Han:2015yys,Cherchiglia:2017uwv,Enomoto:2015wbn}, 
and the model relies on tau-Yukawa to provide the dominant contribution to $(g-2)_\mu$ via two-loop. 
Hence, a light pseudoscalar  in aligned 2HDM will exhibit similar behaviour as in the type-X scenario, 
and one can simply recast our results for this case. 
Alternatively, it is possible to generate substantial $(g-2)_\mu$ at one loop with a $\mathcal{O}(100)$ GeV scalar in muon 
specific~\cite{Abe:2017jqo} or general (type-III) 2HDM framework~\cite{Omura:2015nja}. In muon specific case, 
the BSM scalars couples to muon and the collider signatures are muon rich. Similarly, in 2HDM-III, one has to 
rely on large flavour violating $\mu-\tau$ coupling, which can be probed indirectly via 
$\tau\to\mu\gamma$, $h_{125}\to\mu\tau$ decays. Since the mass scale of the new scalars are generally higher 
than the lepton collider `Higgs factory', these models will show multi muon final state via Yukawa process at 
a high energy lepton collider like CLIC.

\section{Conclusion}\label{sec:conclusion}
The existence of leptophilic spin-0 particles accommodated in the type-X 2HDM is well motivated, particularly,
 in view of  the recent determination of the muon $(g-2)$ excess. 
Search for such a particle at the LHC relies on the exotic decay of the SM Higgs boson to these particles. 
On the other hand, BABAR and beam dump experiments have already excluded them in the sub-GeV mass range. 
The future lepton collider `Higgs factory' will be an ideal place to directly explore GeV scale leptophilic 
bosons and their coupling to leptons. In a previous analysis, we showed that it is possible to 
explore four distinguished tau-lepton signature when the particle mass is few tens of GeV. However, for the 
lighter mass range, this strategy will not work as the decay products of such particle start to merge. 
Here, we propose to probe a muon in association with a charged track as a possible signature for light 
leptophilic spin-0 particles. We have shown that the lepton collider will explore the entire $(g-2)_\mu$ 
allowed parameter space for both scalar and pseudoscalar particles. 
 

\section*{Acknowledgments}
TM would like to thank Akanksha Bhardwaj for useful discussions regarding BDT. TM (EJC) is supported by a KIAS Individual Grant
PG073502 (PG012504) at Korea Institute for Advanced Study.

%
%

\providecommand{\href}[2]{#2}\begingroup\raggedright\endgroup

\end{document}